\documentstyle[preprint,eqsecnum,aps,epsf]{revtex}
\begin{document}
\draft
\title{Level statistics of systems with infinitely many 
independent components based on the Berry-Robnik approach}
\author{Hironori Makino$^1$ and Shuichi Tasaki$^2$}
\address{$^{1}$Department of Human and Information Science, 
Tokai University, C-216, 1117 Kitakaname, Hiratsuka-shi, 
Kanagawa 259-1292, Japan\\
$^{2}$Department of Applied Physics, Waseda University, 3-4-1 
Okubo, Shinjuku-ku, Tokyo 169-8555, Japan}
\date{\today}
\maketitle
\begin{abstract}
Along the line of thoughts of Berry and Robnik{\cite{Ber}}, the limiting gap distribution function of classically integrable quantum systems is derived in the limit of infinitely many independent components.  The limiting gap distribution function is characterized by a single monotonically increasing function $\bar{\mu}(S)$ of the level spacing $S$, and the corresponding level spacing distribution is classified into three cases: (i) Poissonian if $\bar{\mu}(+\infty)=0$, (ii) Poissonian for large $S$, but possibly not for small $S$ if $0<\bar{\mu}(+\infty)< 1$, and (iii) sub-Poissonian if $\bar{\mu}(+\infty)=1$.  This implies that even when the energy-level distributions of individual components are statistically independent, non-Poissonian level spacing distributions are possible.
\end{abstract}
\section{Introduction}
An important property of quantum-classical correspondence appears in the statistical property 
of energy levels of bounded quantum systems in the semiclassical limit.  
Universal behaviors are found in the statistics of {\it unfolded} energy levels at a given 
interval{\cite{Perc,Berr,Bohi}}, which are the sequence of numbers uniquely determined by 
the energy levels using the mean level density obtained from the 
Thomas-Fermi rule{\cite{Boh}}.  It is widely known that, for quantum systems whose classical 
counterparts are integrable (those systems will be referred to as classically integrable systems), 
the distribution of nearest-neighbor level spacing is characterized by the Poisson (exponential) 
distribution{\cite{Berr}}, while for quantum systems whose classical counterparts are strongly 
chaotic, the quantal level statistics are well characterized by the random matrix theory which 
gives level-spacing distribution obeying the Wigner distribution{\cite{Boh,Met}}.  

Level statistics for the integrable quantum systems has been theoretically studied 
by Berry-Tabor{\cite{Berr}}, Sinai{\cite{Sina}}, Molchanov{\cite{Mol}}, Minami{\cite{Minami}}, 
Bleher{\cite{Ble}}, Connors and Keating{\cite{CK}}, and Marklof{\cite{Marklof}}, and have been 
the subject of many numerical investigations.  Still its mechanism is not well understood, the 
appearance of the Poisson distributions is now widely admitted as a universal phenomenon in 
generic integrable quantum systems. 

As suggested, e.g., by Hannay (see the discussion of {\cite{Ber}}), one possible explanation 
would be as follows: For an integrable system of $f$ degrees-of-freedom, almost every orbit 
is generically confined in each inherent torus, and the whole region in the phase space is 
densely covered by invariant tori as suggested by the Liouville-Arnold theorem{\cite{4)}}.  
In other words, the phase space of the integrable system consists of infinitely many tori which 
have infinitesimal volumes in Liouville measure.  Then, the energy level sequence of the whole 
system is a superposition of sub-sequences which are contributed from those regions.  Therefore, 
if the mean level spacing of each independent subset is large, one would expect the Poisson 
distribution as a result of the law of small numbers{\cite{Feller}}.  
This scenario suggested by Hannay is based on the theory proposed by Berry and Robnik{\cite{Ber}}.

The Berry-Robnik theory relates the statistics of the energy level distribution
to the phase-space geometry by assuming that the sequence of the energy spectrum is given by the superposition of statistically independent subspectra, which are contributed respectively from eigenfunctions localized onto the invariant regions
in phase space.  Formation of such independent subspectra is a consequence of the condensation of energy eigenfunctions on disjoint regions in the classical phase space and of the lack of mutual overlap between their eigenfunctions, and, thus, can only be expected in the semi-classical limit where the Planck constant tends to zero, $\hbar\to 0$.  This mechanism is sometimes referred to as {\it{the principle of uniform semi-classical condensation of eigenstates}}{\cite{Berr77,Robn}}.  This principle states that the Wigner function of a semiclassical eigenstate is connected on a region in phase space explored by a typical trajectory of the classical dynamical system.  In integrable systems, the phase space is folded into invariant tori, and 
the Wigner functions of the eigenstates tend to delta functions on these tori in the semiclassical limit{\cite{Berr77-2}}.  On the other hand, in a strongly chaotic system, almost all trajectories cover the energy shell uniformly, and hence the Wigner 
functions of eigenstates are expected to become a delta function on the energy shell as suggested by the quantum ergodicity theorem{\cite{erq1,erq2}}.  Because of the suppression of the tunneling, each quantum eigenstate is folded into independent subsets in the semiclassical limit, and is expected to form independent spectral components.  Indeed, formation of such independent components are checked numerically in a deep semi-classical regime{\cite{regime}}.

In the Berry-Robnik approach{\cite{Ber}}, the overall level spacing distribution is derived along a line of mathematical framework by Mehta{\cite{Met}}, as follows: Consider a system whose classical phase space is decomposed into $N$-disjoint regions.  The Liouville measures of these regions are denoted by $\rho_i(i=1,2,3,\cdots,N)$ which satisfy $\sum_{i=1}^N\rho_i= 1$.  Let $E(S)$ be the gap distribution function which stands for the probability that an interval $(0,S)$ contains no level.  $E(S)$ is expressed by the level spacing distribution $P(S)$ as follows;
%
\begin{equation}
E(S)=\int_S^\infty d\sigma \int_\sigma^\infty P(x)dx.
\label{eq:1-1}
\end{equation}
%
When the entire sequence of energy levels is a product of the statistically independent superposition of $N$ sub-sequences, $E(S;N)$ is decomposed into those of sub-sequences, $E_i(S;\rho_i)$,
%
\begin{equation}
E(S;N)=\prod_{i=1}^N E_i(S;\rho_i).
\label{eq:1-2}
\end{equation}
%
In terms of the normalized level spacing distribution $p_i(S;\rho_i)$ of a sub-sequence, $E_i(S;\rho_i)$ is given by 
%
\begin{equation}
E_i(S;\rho_i)= \rho_i\int_S^\infty d\sigma 
\int_\sigma^\infty p_i(x;\rho_i)dx,
\label{eq:1-3}
\end{equation}
%
and $p_i(S;\rho_i)$ is assumed to satisfy{\cite{Ber}}
%
\begin{equation}
\int_0^{\infty} S\cdot p_i(S;\rho_i)dS =\frac{1}{\rho_i}.
\label{eq:1-4}
\end{equation}
%
Equations ({\ref{eq:1-2}}) and ({\ref{eq:1-4}}) relate the level statistics in the semiclassical limit with the phase-space geometry. 

In most general cases, the level spacing distribution might be singular.  In such a case, it is convenient to use its cumulative distribution function $\mu_i$;
\begin{equation}
\mu_i(S)=\int_0^S p_i(x;\rho_i)dx.
\label{eq:1-5}
\end{equation}

In addition to equations ({\ref{eq:1-2}}) and ({\ref{eq:1-4}}), we assume two conditions for the statistical weights: 
\begin{itemize}
\item Assumption (i): The statistical weights of independent regions uniformly vanishes in the limit of infinitely many regions;
\begin{equation}
\max_i \rho_i \to 0\quad\mbox{as}\quad N\to +\infty.
\label{eq:1-7}
\end{equation}
\item Assumption (ii): The weighted mean of the cumulative distribution of energy spacing,
\begin{equation}
\mu(\rho;N)=\sum_{i=1}^N \rho_i\mu_i(\rho),
\label{eq:1-8}
\end{equation}
converges in $N\to +\infty$ to $\bar{\mu}(\rho)$
\begin{equation}
\lim_{N\to +\infty}\mu(\rho;N) = \bar{\mu}(\rho).
\label{eq:1-9}
\end{equation}
The limit is uniform on each closed interval: \ $0\le \rho \le S$.
\end{itemize}
In the Berry-Robnik theory, the statistical weights of individual components are related to the phase volumes of the corresponding invariant regions.  This relation is satisfactory if the Thomas-Fermi rule for the individual phase space regions still holds, and thus supports eq.({\ref{eq:1-4}}){\cite{Ber}}.  Here we do not specify the validity of this problem and deal with the statistical weights as parameters.

Under assumptions (i) and (ii), eqs.({\ref{eq:1-2}}) and ({\ref{eq:1-4}}) lead to the overall level spacing distribution whose gap distribution function is given by the following formula in the limit of $N\to +\infty$,
%
\begin{equation}
E_{\bar{\mu}}(S)=
\exp{\left[-\int_0^S
\left(1-\bar{\mu}(\sigma)\right)d\sigma \right]}
\label{eq:1-10},
\end{equation}
%
where the convergence is in the sense of the weak limit.  When the level spacing distributions of individual components are sparse enough, one may expect $\bar{\mu}=0$ and the level spacing distribution of the whole energy sequence reduces to the Poisson distribution,
%
\begin{equation}
E_{\bar{\mu}=0}(S)=\exp{\left(-S\right)}.
\label{eq:1-11}
\end{equation}
%
%
In general, one may expect $\bar{\mu}\not=0$ which corresponds to a certain accumulation of the levels of individual components. 

In the following sections, the above statement is proved and the limiting level spacing distributions are classified into three classes.  One of them is the Poisson distribution as discussed in the original work by Berry and Robnik{\cite{Ber}}.
The others are not Poissonian.  We give examples of non-Poissonian limiting level spacing distributions in section {\ref{secIII}}.  In the concluding section, we discuss some relations between our results and other related works.
\section{Limiting level spacing Distribution}
\label{secII}
\subsection{Derivation of the limiting Gap distribution}
In this section, starting from eqs.({\ref{eq:1-2}}) and ({\ref{eq:1-4}}), and assumptions (i) and (ii) introduced in the previous section, we show that, in the limit of infinitely many components $N\to +\infty$, the gap distribution $E(S;N)$ converges to the distribution function ({\ref{eq:1-10}}) with $\bar{\mu}$.  
The convergence is shown as follows.

Following the procedure by Mehta(see appendix A.2 of Ref.{\cite{Met}}), we rewrite $E(S;N)$ in terms of the cumulative level spacing distribution $\mu_i(S)$ 
of independent components:
\begin{equation}
E(S;N)=\prod_{i=1}^N \left[\rho_i \int_S^{+\infty} 
d\sigma \left(1-\mu_i(\sigma)\right)\right].
\label{001}
\end{equation}
Equation (1.4) and integration by parts lead to
$$
\int_0^{+\infty}d\sigma (1-\mu_i(\sigma))=\sigma(1-\mu_i(\sigma))|_0^{+\infty}
+\int_0^{+\infty}\sigma p_i(\sigma)d\sigma =\frac{1}{\rho_i},
$$
where $\lim_{\sigma\to+\infty}\sigma(1-\mu_i(\sigma))=0$ follows from 
the existance of average, and hence to
\begin{equation}
\rho_i\int_S^{+\infty}d\sigma(1-\mu_i(\sigma))
=1-\rho_i\int_0^S d\sigma(1-\mu_i(\sigma)).
\label{002}
\end{equation}
Since the convergence of 
$\sum_{i=1}^N \rho_i \mu_i(\sigma) \to {\bar \mu}(\sigma)$ 
for $N\to +\infty$ is uniform on each 
interval $\sigma \in [0,S]$ by Assumption (ii), $E(S;N)$ has 
the following limit,
\begin{eqnarray}
\log{E(S;N)}&=&\sum_{i=1}^N \log{\left[1-\rho_i\int_0^{S}d\sigma(1-\mu_i(\sigma))\right]}\nonumber\\
&=&-\sum_{i=1}^N\left[\rho_i\int_0^{S}d\sigma
(1-\mu_i(\sigma))+O(\rho_i^2)\right]\nonumber\\
&=&-\int_0^S d\sigma \left[1-\mu(\sigma;N)\right]
+\sum_i^N O(\rho_i^2)\label{eq:sum}\\
&&\longrightarrow - \int_0^S d\sigma 
\left[1-\bar{\mu}(\sigma)\right]
\quad\mbox{as}\quad N\to +\infty \label{lim:01},
\end{eqnarray}
where we have used $|\mu_i(\sigma)|\le 1$, 
$\log(1+\epsilon)=\epsilon+O(\epsilon^2)$ 
in $\epsilon\ll 1$, and the following 
property obtained from Assumption (i),
\begin{equation}
|\sum_{i=1}^N O(\rho_i^2)|\leq C\cdot
\max_i{\rho_i}\cdot\sum_{i=1}^N 
\rho_i =C\cdot\max_i{\rho_i}\to 0
\quad\mbox{as}\quad N\to+\infty,
\label{lim:O}
\end{equation}
with $C$ a positive constant.  
Therefore, we have the desired result:
\begin{equation}
\lim_{N\to+\infty}E(S;N)=E_{\bar{\mu}}(S)
=\exp{\left[-\int_0^S
\left(1-\bar{\mu}(\sigma)\right)d\sigma \right]}.
\label{eq:2-7}
\end{equation}
\subsection{Weak convergence limit of the level spacing distribution}
In this section, we show that in $N\to+\infty$ limit, the level spacing 
distribution $P(S;N)$ converges weakly to $P_{\bar{\mu}}(S)$,
\begin{equation}
P_{\bar{\mu}}(S)=\frac{d^2}{dS^2}E_{\bar{\mu}}(S).
\end{equation}

According to the Helly's theorem{\cite{Sin,Feller}}, the weak convergence of the level spacing distribution is defined by
\begin{equation}
\lim_{N\to+\infty}\int_0^S P(x;N)dx = \int_0^S P_{\bar{\mu}}(x)dx.
\label{aaa}
\end{equation}
Since each side of the above equation is rewritten as,
$$
\int_0^S P(x;N)dx = \left[\frac{d}{dx}E(x;N)\right]_0^S,\ 
\int_0^S P_{\bar{\mu}}(x)dx = 
\left[\frac{d}{dx}E_{\bar{\mu}}(x)\right]_0^S,
$$
the limit ({\ref{aaa}}) is equivalent to
$$
\lim_{N\to+\infty}\frac{d}{dS}E(S;N)=\frac{d}{dS}E_{\bar{\mu}}(S).
$$
The above equation is proved as follows: By using equation ({\ref{001}}), we rewrite $\frac{d}{dS}E(S;N)$ in terms of the cumulative level spacing distribution function $\mu_i(S)$ of spectral components,
$$
\frac{d}{dS}E(S;N)=-E(S;N)\sum_{i=1}^N {1- \mu_i(S)\over 
\rho_i\int_{S}^{+\infty}d\sigma (1-\mu_i(\sigma))}.
$$
In the limit $N\to+\infty$, one has 
$E(S;N)\to E_{\bar{\mu}}(S)$ as shown by equation ({\ref{eq:2-7}}) and,
\begin{eqnarray}
\sum_{i=1}^N {1-\mu_i(S)\over 
\rho_i\int_S^{+\infty}d\sigma (1-\mu_i(\sigma))}
&=&\sum_{i=1}^N {\rho_i-\rho_i \mu_i(S)\over 1-
\rho_i\int_0^{S}d\sigma (1-\mu_i(\sigma))}\nonumber\\
&=& 1-\sum_{i=1}^N\rho_i\mu_i(S)
+\sum_{i=1}^N O(\rho_i^2)\nonumber\\
&&\longrightarrow 1-\bar{\mu}(S)
\quad\mbox{as}\quad N\to +\infty,\nonumber
\end{eqnarray}
where we have used equation ({\ref{002}}), 
$1/(1-\epsilon)=1+O(\epsilon)$ in $\epsilon\ll 1$, 
and the limit ({\ref{lim:O}}).  Therefore, we have 
the desired result:
\begin{equation}
\lim_{N\to+\infty}\frac{d}{dS}E(S;N)
=-\left( 1-\bar{\mu}(S) \right) E_{\bar{\mu}}(S)
=\frac{d}{dS}E_{\bar{\mu}}(S).
\label{RRR}
\end{equation}

We remark that, when the limiting distribution function 
${\bar \mu}(S)$ is differentiable, the asymptotic level spacing 
distribution is described as follows: 
\begin{equation}
P_{\bar{\mu}}(S)= \left[(1-\bar{\mu}(S))^2 
+ \bar{\mu}'(S) \right] \exp{\left[-\int_0^S
\left(1-\bar{\mu}(\sigma)\right)d\sigma \right]}
\label{Density}.
\end{equation}
\subsection{Properties of the limiting level spacing distribution}
Since $\mu_i(S)$ is monotonically increasing 
and $0 \le \mu_i(S) \le 1$, 
$\bar{\mu}(S)$ has the same properties. 
Then, $1-\bar{\mu}(S)\ge 0$ for any
$S\ge 0$ and one has
\begin{equation}
\frac{1}{S}\int_0^S d\sigma 
(1-\bar{\mu}(\sigma))\longrightarrow 
1-\bar{\mu}(+\infty)
\quad\mbox{as}\quad S\to+\infty.
\label{eq:4-7}
\end{equation}
According to the above limit, the level spacing distribution is classified into the following three cases in the sense of weak limit:
\begin{itemize}
\item Case 1,~$\bar{\mu}(+\infty)=0$: The limiting level spacing distribution is the Poisson distribution.  Note that this condition is equivalent to $\bar{\mu}(S)=0$ for ${}^{\forall} S$ because $\bar{\mu}(S)$ is monotonically increasing.
\item Case 2,~$0<\bar{\mu}(+\infty)<1$: 
For large $S$ values, the limiting level spacing distribution is well approximated by the Poisson distribution, while, for small $S$ values, it may deviate from the Poisson distribution.
\item Case 3,~$\bar{\mu}(+\infty)=1$: 
The limiting level spacing distribution deviates from the Poisson distribution for ${}^{\forall}S$, and decays as $S\to+\infty$ more slowly than does the Poisson distribution.  This case will be referred to as a sub-Poisson distribution. 
\end{itemize}
One has Case 1 if the individual cumulative distribution function $\mu_i(S)$ are bounded by a finite positive function $g(S)$, 
$$
\mu_i(S)\leq \rho_i^{\eta} g(S),
$$
for ${}^{\forall}i$ 
and $\eta >0$.  Indeed, one has,
$$
\mu(S;N)=\sum_{i=1}^N \rho_i\mu_i(S)\leq 
g(S)\sum_i^N \rho_i\rho_i^{\eta}\leq g(S)(\max_i\rho_i)^{\eta}
\longrightarrow 0\equiv\bar{\mu}(S),\quad\mbox{as}\quad N\to\infty.
$$
More specifically, for example, one has Case 1 if the individual level spacing distributions are derived from scaled distribution functions $f_i$ as{\footnote{Eq.({\ref{eq:1-4}}) is described by rewriting $x=\rho_i S$ in the following way, $1=\int S\rho_i p_i(S;\rho_i)dS=\int x f_i(x)dx$, so that $f_i(x)=\rho_i p_i(x;\rho_i)$ is a dimensionless function.  For instance, $f_i(x)=\exp{(-x)}$ in the case of Poisson distribution $p_i(S;\rho_i)=\rho_i
\exp{\left(-\rho_i S\right)}$, and $f(x)=
\frac{\pi}{2}x\exp{\left(-\frac{\pi}{4}x^2\right)}$ in the case 
of the Wigner distribution $p_i(S;\rho_i)=
\frac{\pi}{2}\rho_i^2 S
\exp{\left[-\frac{\pi}{4}\rho_i^2 
S^2\right]}$.}},
$$
p_i(S;\rho_i)=\rho_i f_i(\rho_i S),
$$
where $f_i$ satisfy
$$
\int_0^{+\infty} f_i(x) dx = 1 , 
\qquad \int_0^{+\infty} x f_i(x) dx = 1,
$$
and are uniformly bounded by a 
positive constant $D$: $|f_i(S)|\le D$
($1\le i \le N$ and $S\ge 0$). 
Indeed, one then has
\begin{eqnarray}
|\mu(S;N)|
&\leq& 
\left|\sum_{i=1}^N\rho_i \int_0^S p_i(x,\rho_i)dx\right|\\
&\leq& \sum_{i=1}^N \rho_i^2\int_0^S
\left| f_i\left({\rho_i}x\right)\right|dx \nonumber \\
&\leq& DS\sum_{i=1}^N\rho_i^2
\leq DS\max_i\rho_i\sum_{i=1}^N\rho_i
\longrightarrow 0\equiv {\bar \mu}(S).
\end{eqnarray}

This includes the case studied by Berry and Robnik{\cite{Ber}}, where the gap distribution is a product of superposition of a single regular component characterized by the Poisson distribution and $N$ equivalent chaotic components characterized by the Wigner distribution, and the latter is expressed by the product of the scaled distributions as:
\begin{equation}
E^{BR}(S;N)=\exp{\left(-\rho_0 S\right)}
\prod_{i=1}^N E_i^{\mbox{\tiny WIGNER}}(S;\rho_i),
\label{eq2-12}
\end{equation}
where the statistical weights are $\rho_i = {1-\rho_0\over N}$
and the individual level spacing distributions $f_i$ 
corresponding to the gap distributions 
$E_i^{\mbox{\tiny WIGNER}}(S;\rho_i)=
\mbox{erfc}\left(\frac{\sqrt{\pi}}{2}\rho_i S\right)$ 
are given by the dimensionless Wigner distribution:
\begin{equation}
f_i(x)= \frac{\pi x}{2} 
\exp{\left[-\frac{\pi}{4}x^2 
\right]}.
\end{equation}
Indeed, one has the Poisson distribution in $N\to+\infty$ limit:
\begin{eqnarray}
E^{\mbox{\tiny{BR}}}(S;N)&=&
\exp{\left[-\rho_0 S 
+(N-1)\log{\mbox{erfc}\left(\frac{\sqrt{\pi}}{2}
\frac{1-\rho_0}{N-1}S\right)}
\right]}\longrightarrow e^{-S}.
\end{eqnarray}
\section{Example}
\label{secIII}
As an example of the deviation from the Poisson 
distribution, we study rectangular billiard system whose 
energy levels are described by using positive integer numbers, 
$m$ and $i$ as follows,
\begin{equation}
\epsilon_{m,i} =  m^2 + \alpha\  i^2,
\label{eq:e}
\end{equation}
where $\alpha$ is the aspect ratio of two sides of a billiard 
wall.  For a given energy 
interval $[\epsilon,\epsilon+\Delta \epsilon]$, each 
energy level is classified into 
components according 
to the eigenvalues, $i=1,2,3,\cdots,N$;
\begin{equation}
N= \left[ \sqrt{\frac{1+\gamma}{\alpha}\epsilon\mbox{ }}\right],
\label{eq:cx}
\end{equation}
where $\gamma\equiv\Delta\epsilon/\epsilon$, and $[x]$ 
stands for the maximum integer which does not exceed $x$.  
The relative weight of each component is given by
\begin{equation}
\rho_i=\left\{
\begin{array}{lr}
\frac{4(1+\gamma)}{N\pi\gamma}
\left(\sqrt{1-\left(\frac{i}{N}\right)^2}-
\sqrt{\frac{1}{1+\gamma}-\left(\frac{i}{N}\right)^2}\right)+O
\left(\frac{1}{N^2}\right)
\quad&\mbox{if}\quad i <  \frac{N}{\sqrt{1+\gamma}},
\nonumber\\
\frac{4(1+\gamma)}{N\pi\gamma}
\left(\sqrt{1-\left(\frac{i}{N}\right)^2}
\right)+O\left(\frac{1}{N^2}\right)
\quad&\mbox{if}\quad \frac{N}{\sqrt{1+\gamma}}\leq i\leq N.
\end{array}\right.
\end{equation}
As easily seen, $\rho_i$ satisfies 
the assumption (i);
\begin{equation}
\max_i \rho_i \leq \frac{4}{N\pi}
\sqrt{1+\frac{1}{\gamma}}+O\left(\frac{1}{N^2}\right)
\longrightarrow 0 \qquad 
\mbox{as}\quad N\to+\infty.
\end{equation}
Note that the limit of infinitely many components, 
$N\to+\infty$, corresponds 
to the high energy limit, 
$\epsilon\to+\infty$(see Eq.({\ref{eq:cx}})), which 
is equivalent to the semiclassical limit in physical systems.  In this 
limit, the statistical weight of each sub-spectrium becomes 
sparse, since each element of $\mu(S;N)$, $\rho_i\mu_i(S)$, 
tends to zero: $\rho_i\mu_i(S)\leq\max_j\rho_j\to0$.

Figures 1(a) and 1(b) show numerical results of the level-spacing distribution $P(S)$ for two values of $\alpha$, and figures 2(a) and 2(b) show the gap distribution function corresponding to figures 1(a) and 1(b), respectively.  In case that $\alpha$ is far from rational, $P(S)$ and $E(S)$ are well approximated by the Poisson distribution while in case that $\alpha$ is close to a rational expressed as $\alpha=p/q$, where $p$ and $q$ are coprime positive integers, they deviate from the Poisson distribution\cite{Rob}.

In order to compare the non-Poisson distribution and the classification given in the previous section, we consider,
\begin{equation}
\tilde{\mu}(S;N)=1-\frac{1-\int_0^S P(x;N)dx}{E(S;N)}.
\end{equation}
When $N\to+\infty$, $\tilde{\mu}(S;N)$ approaches 
$\bar{\mu}(S)$, and this function distinguishes the 
three cases as follows: In Case 1 i.e., where the level spacing 
obeys the Poisson distribution, $\lim_{N\to+\infty}\tilde{\mu}(S;N)=0$.  
In Case 2, $\lim_{N\to+\infty}\tilde{\mu}(S;N)\to c$ as $S\to+\infty$ 
$(0<c<1)$, and in Case 3, where the sub-Poisson distribution is expected, 
$\lim_{N\to+\infty}\tilde{\mu}(S;N)\to1$ as $S\to+\infty$.

Figure 3 shows $\tilde{\mu}(S;N)$ for different 
values of $N$.  The dotted line $\tilde{\mu}=0$ exhibits the 
Poisson distribution.  From this, one can think that $\tilde{\mu}(S;N)$ for 
$N=61905,S\leq 10$ well approximates $\lim_{N\to+\infty}\tilde{\mu}(S;N)$. 

Figure 4 shows $\tilde{\mu}(S;N)$ for the two values of 
$\alpha$ corresponding to figures 2(a)--2(b), respectively.  
In case that the numerical data is well characterized by the 
Poisson distribution (figures 1(a) and 2(a)), the corresponding 
function $\tilde{\mu}(S;N)$ agrees with $0$, while in case that 
deviates from the Poisson distribution (figures 1(b) and 2(b)), 
$\tilde{\mu}(S;N)\ne0$ and $\tilde{\mu}(S;N)\to c(0<c<1)$ 
for $S\to+\infty$.  Therefore, this result corresponds to the 
Case 2.  

In this model, we have not yet observed 
the clear evidence of Case 3.   Such a case is expected when 
there is stronger accumulation of the energy 
levels of individual components.
\section{Extended formalism of the Berry-Robnik distribution}
\label{extension}
In this section, we propose one possible extension of the 
Berry-Robnik distribution (\ref{eq2-12}) for the level statistics 
of the nearly-integrable system with two degree-of-freedom. 
Since the classical phase space of the nearly-integrable system 
consists of regular and chaotic regions and the Liouville 
measures of the chaotic regions are larger than zero, 
$\rho_i>0$, this system does not support the assumption (i). 
However, the regular regions consist of infinitely 
many subsets, and our approach shown in section {\ref{secII}} 
is partially applicable to the spectral components 
corresponding to the regular regions.

Following to the assumption proposed by Berry and Robnik{\cite{Ber}}, 
the gap distribution functions in the nearly integrable system, 
which are contributed from the individual chaotic 
regions, are characterized by the Random Matrix Theory(RMT).  
Then one has,
\begin{eqnarray}
\log E(S;N_1;N_2) &=&\log 
\prod_{i=1}^{N_1}E_i^{\mbox{\tiny{RMT}}}(S;\rho_i) 
\prod_{j=1}^{N_2}E_j(S;\rho_j^{'})\\
&=&\sum_{i=1}^{N_1}\log E_i^{\mbox{\tiny{RMT}}}(S;\rho_i)
+\sum_{j=1}^{N_2}\log E_j (S;\rho_j^{'}),
\end{eqnarray}
where $\rho_j^{'}=\rho_{j+N_1}$, and 
$E_j(S;\rho_j^{'})$ denote the gap distribution functions 
corresponding to the subsets in the regular regions.   
As shown by eq.(2.6), $E_j$ has the following limit,
$$
\sum_{j=1}^{N_2} \log E_j (S;\rho_j^{'}) \longrightarrow \rho_0\int_0^S 
d\sigma\left[1-\bar{\mu}(\sigma)\right]\ \mbox{as}\ N_2\to+\infty,
$$
where $\rho_0=\sum_{j=1}^{N_2}\rho_j^{'}$ and $\bar{\mu}(\sigma)=
\lim_{N_2\to+\infty}\sum_{j=1}^{N_2}\rho_j^{'}\mu_j(\sigma)$.
Accordingly, in the partial limit of $N_1\ll+\infty, N_2\to+\infty$, 
the original proposal for the gap distribution by Berry and Robnik 
is replaced by
\begin{equation}
\lim_{N_2\to+\infty}E(S;N_1;N_2)=E_{\bar{\mu}}(S;N_1)=
\exp{\left[-\rho_0 \int_0^S(1-\bar{\mu}(\sigma))
d\sigma\right]}\prod_{i=1}^{N_1} E_i^{\mbox{\tiny RMT}}(S;\rho_i),
\label{nf1}
\end{equation}
where $\rho_0$ 
denotes the total amount of the Liouville measures 
of the regular region in mixed phase space.  The above 
distribution formula is classified into the following three 
cases;  Case 1', $\bar{\mu}(+\infty)=0$: Berry-Robnik distribution,  
Case 2', $0<\bar{\mu}(+\infty)< 1$: Berry-Robnik 
distribution for large $S$, 
but possibly not for small $S$, 
and Case 3', $\bar{\mu}(+\infty)=1$:  A distribution 
function obtained by the superposition of spectral 
components obeying the sub-Poisson 
statistics and the Random matrix theory.  From this 
classification, one 
can see that the new formula ({\ref{nf1}}) admits 
deviations from the Berry-Robnik distribution 
when $\bar{\mu}(+\infty)\ne 0$.  

We remark that $P(S;N_1,N_2)=\frac{d^2}{dS^2}E(S;N_1,N_2)$ in the limit
$N_2\to+\infty$ converges 
weakly to the limiting level spacing distribution: 
$P_{\bar{\mu}}(S;N_1)=\frac{d^2}{dS^2}E_{\bar{\mu}}(S;N_1)$, and 
when the limiting function ${\bar \mu}(S)$ is differentiable, 
the asymptotic level spacing distribution admits the 
following density: 
\begin{equation}
P_{\bar{\mu}}(S;N_1)=\frac{d^2}{dS^2}\left[
\exp{\left(-\rho_0 \int_0^S(1-\bar{\mu}(\sigma))
d\sigma\right)}\prod_{i=1}^{N_1} E_i^{\mbox{\tiny RMT}}(S;\rho_i)\right]
\label{nf2}
\end{equation}
\indent
The validity of the Berry-Robnik distribution has been checked 
for generic nearly-integrable systems 
by many numerical 
investigations{\cite{fract1,fract2,Li,Pro,makino1,makino2,makino3}}.  
Among them, Prosen and Robnik found numerically for several systems 
that there is a high energy region in which the Berry-Robnik 
distribution formula (\ref{eq2-12}) well approximates the 
level spacing distribution{\cite{fract1}}.  This energy region 
is sometimes referred to as {\it the Berry-Robnik regime}{\cite{regime}}.  
While they also found that the level spacing distribution in the low energy region deviates from the Berry-Robnik formula, and approximates 
the Brody distribution.  This behavior was studied 
in terms of a fractional power dependence 
of the spacing distribution 
near the origin at $S=0$, which could 
be attributed to the localization 
properties of eigenstates on chaotic components{\cite{fract1,fract2}}.  
From the above classification, Case 1':$\bar{\mu}(+\infty)=0$ sould 
be satisfied in the Berry-Robnik regime.  While Case 2' and Case 3' 
might propose another possibilities.  
When the spectral components corresponding to regular regions show 
strong accumulation, the gap distribution obeys 
the distribution formula ({\ref{nf1}}) with $0<\bar{\mu}(+\infty)\leq 1$, 
and the level spacing distribution shows deviations from 
the Berry-Robnik distribution.  Therefore, this result 
might propose another possibility of the Berry-Robnik 
approach.
%
%
%
%
\section{Conclusion and Discussion}
In this paper, we investigated the gap distribution 
function of systems with infinitely many independent 
components and discussed the level-spacing 
statistics of classically integrable quantum systems.  
In the semiclassical limit, reflecting infinitely fine classical 
phase space structures, individual energy eigenfunctions are expected 
to be well localized in the phase space and contribute independently 
to the level statistics.  Keeping this expectation in 
mind, we considered a situation in which the system consists of 
infinitely many components and each of them gives an infinitesimal 
contribution.  And by applying the arguments of Mehta, and Berry 
and Robnik, the limiting level spacing distribution 
was obtained whose gap distribution function is described by 
a single monotonically increasing function 
$\bar{\mu}(S)$ of the level spacing $S$:
\begin{equation}
E_{\bar{\mu}}(S)= \exp{\left[-\int_0^S
\left(1-\bar{\mu}(\sigma)\right)d\sigma \right]}
\label{eq:XS}
\end{equation}
The weak convergence limit 
of the level spacing distribution is classified into 
three cases; Case 1: Poissonian if 
$\bar{\mu}(+\infty)=0$, Case 2: Poissonian 
for large $S$, but possibly not for small 
$S$ if $0<\bar{\mu}(+\infty)< 1$, 
and Case 3: sub-Poissonian if $\bar{\mu}(+\infty)=1$. 
Thus, even when the 
energy levels of individual components are statistically 
independent, non-Poissonian level spacing distributions 
are possible.

In most general cases, the integral in equation ({\ref{eq:XS}}) 
converges in $S\ll+\infty$ and then 
$\lim_{S\to+\infty}E_{\bar{\mu}}(S)\not=0$, 
the limiting gap distribution $E_{\bar{\mu}}(S)$ does 
not work accurately.  In such case, however, its 
differentiation ({\ref{RRR}}) still work accurately 
in $S\to+\infty$ limit{\cite{MAK_TAS}}, 
and thus the above classification (Case 1--3) 
holds in general.

Note that the singular level spacing distribution can be taken into 
account in terms of non-smooth cumulative 
distributions.  Such a singularity is expected when there is strong 
accumulation of the energy levels of individual components.  For a 
certain class of systems, such accumulation is observable.  
One example is shown in section III where the results show clear 
evidence of Case 2.  Another example is the two-dimensional 
harmonic oscillator whose 
level spacing distribution is non-smooth for arbitrary 
system parameter\cite{Berr,Ble}.  The final example is studied by 
Shnirelman\cite{Shni}, Chirikov and Shepelyansky\cite{Shni2}, 
and Frahm and Shepelyansky\cite{Frahm} for a certain type of 
system which contains a quasi-degeneracy result from 
inherent symmetry(time reversibility).  
As is well known, the existence of quasi-degeneracy leads 
to the sharp Shnirelman peak at small spacings.

Finally, in section \ref{extension}, we proposed one possible extension 
of the Berry-Robnik distribution for classically 
nearly-integrable quantum systems.  This extension admitted 
deviations from the Berry-Robnik distribution 
when there is strong accumulation of the energy levels of 
spectral components.  Such possibilities will be studied 
elsewhere.
\vspace{0.2cm}\\
\vspace{0.2cm}\\
{\large Acknowledgments}\\
\indent
The authors would like to thank Professor M. Robnik, Professor 
Y.Aizawa, Professor A.Shudo, and Dr G. Veble for their 
helpful advices and discussions.  The authors also thank the 
Yukawa Institute for Theoretical Physics at Kyoto University. 
Discussions during the YITP workshop YITP-W-02-13 on "Quantum chaos: 
Present status of theory and experiment" were useful to 
complete this work.   This work is partly supported by 
a Grant-in-Aid for Scientific Research 
(C) from the Japan Society for the 
Promotion of Science. 

\newpage
\begin{figure}
\epsfxsize = 12cm
\centerline{\epsfbox{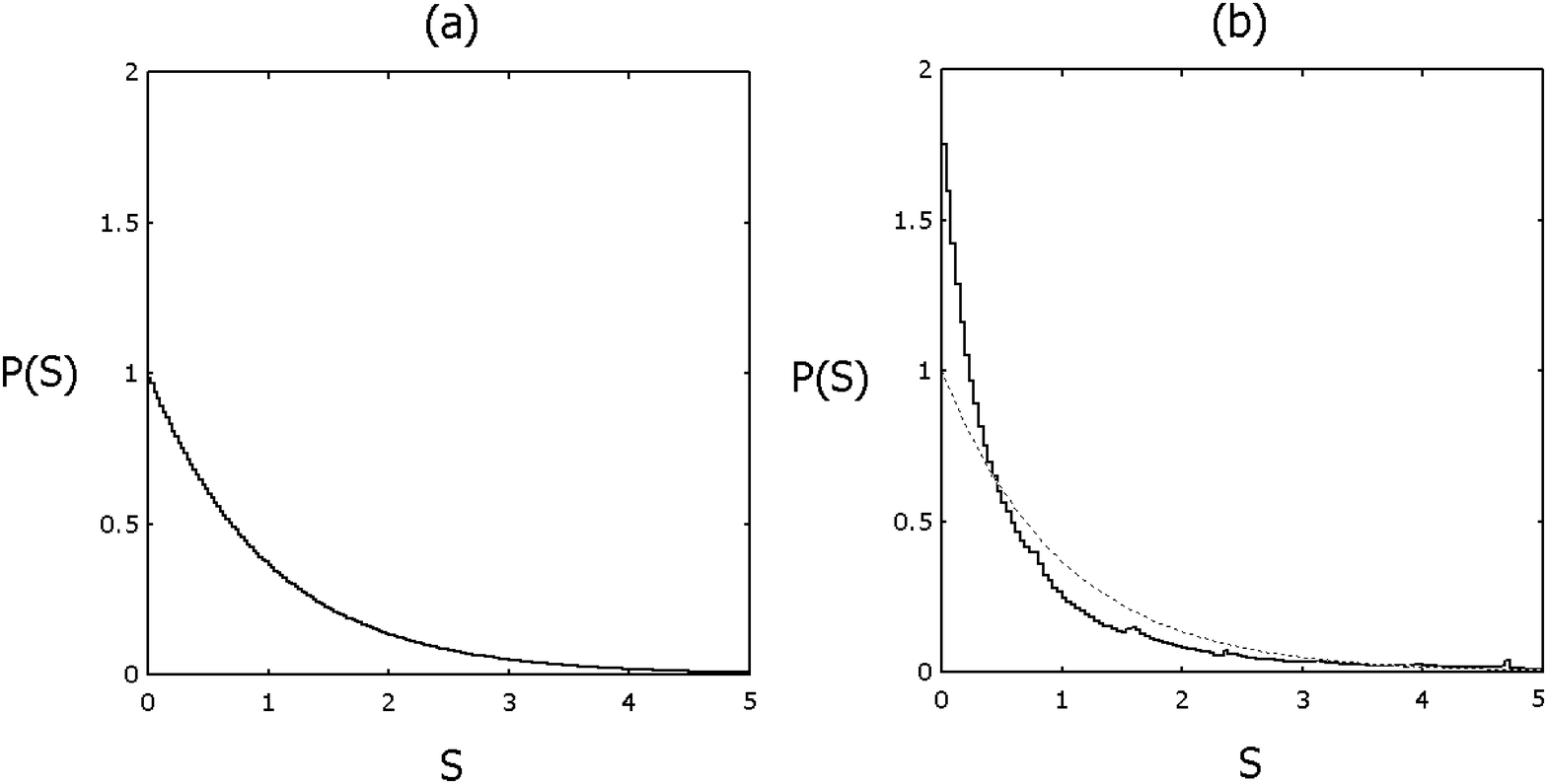}}
{FIG.1  Numerical results of the level spacing 
distribution $P(S)$ for 
(a) $\alpha=1+\frac{\pi}{3}\times10^{-4}$,
(b) $\alpha=1+\frac{\pi}{2}\times10^{-9}$.
We used energy levels $\epsilon_{m,l}$ with 
$\frac{\pi}{4\sqrt{\alpha}}\epsilon_{m,l} 
\in [300\times10^7,301\times10^7]$.  
Total numbers of levels are (a) 10000016, 
(b) 10000046.  The dotted 
curve in each figure shows 
the Poisson distribution: $P(S)=e^{-S}$.}
\label{fig_1}
\end{figure}
\begin{figure}
\epsfxsize = 12cm
\centerline{\epsfbox{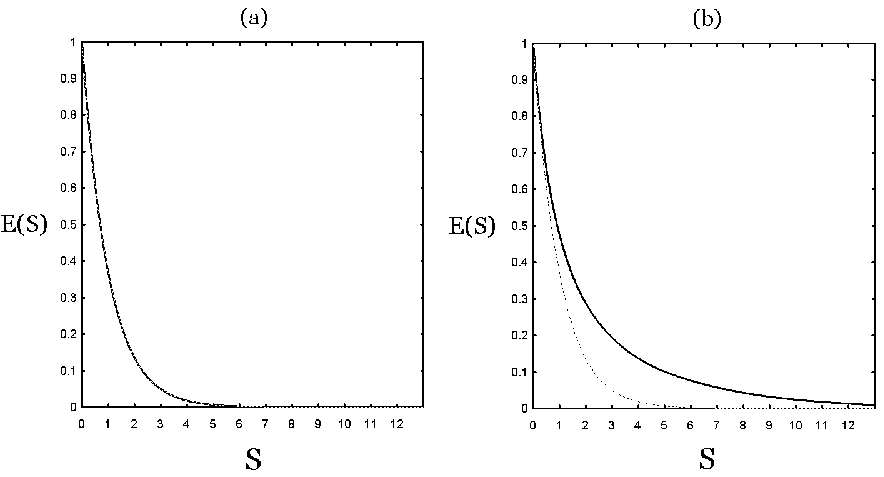}}
{FIG.2  The gap distribution function $E(S)$ for 
(a) $\alpha=1+\frac{\pi}{3}\times10^{-4}$, 
(b) $\alpha=1+\frac{\pi}{2}\times10^{-9}$.  
The dotted curve in each figure exhibits the Poisson 
distribution: $E(S)=e^{-S}$.}
\label{fig_2}
\end{figure}
\begin{figure}
\epsfxsize = 8cm
\centerline{\epsfbox{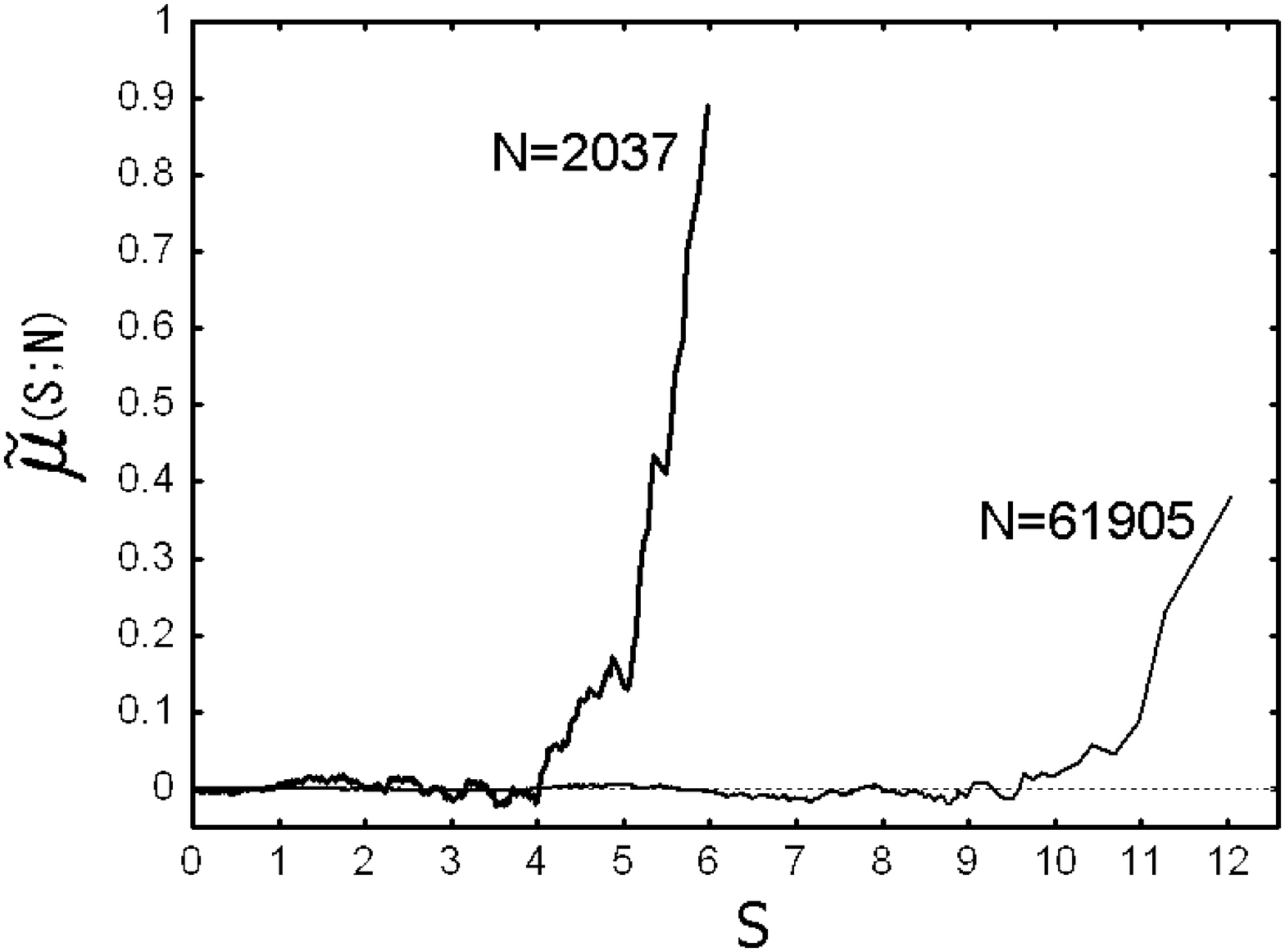}}
{FIG.3  $\tilde{\mu}(S;N)$ for $N=2037$ and for $N=61905$. 
In each case, we fixed $\alpha=1+\frac{\pi}{3}\times 10^{-4}$, and used 
$30048$ energy levels with $\frac{\pi}{4\sqrt{\alpha}}\epsilon_{m,l} 
\in [323\times10^4,326\times10^4]$ 
and $10000016$ energy levels with $\frac{\pi}{4\sqrt{\alpha}}\epsilon_{m,l} 
\in [300\times10^7,301\times10^7]$, respectively.  The dashed 
line ($\tilde{\mu}=0$) exhibits the Poisson distribution.}
\label{fig_4}
\end{figure}
\begin{figure}
\epsfxsize = 8cm
\centerline{\epsfbox{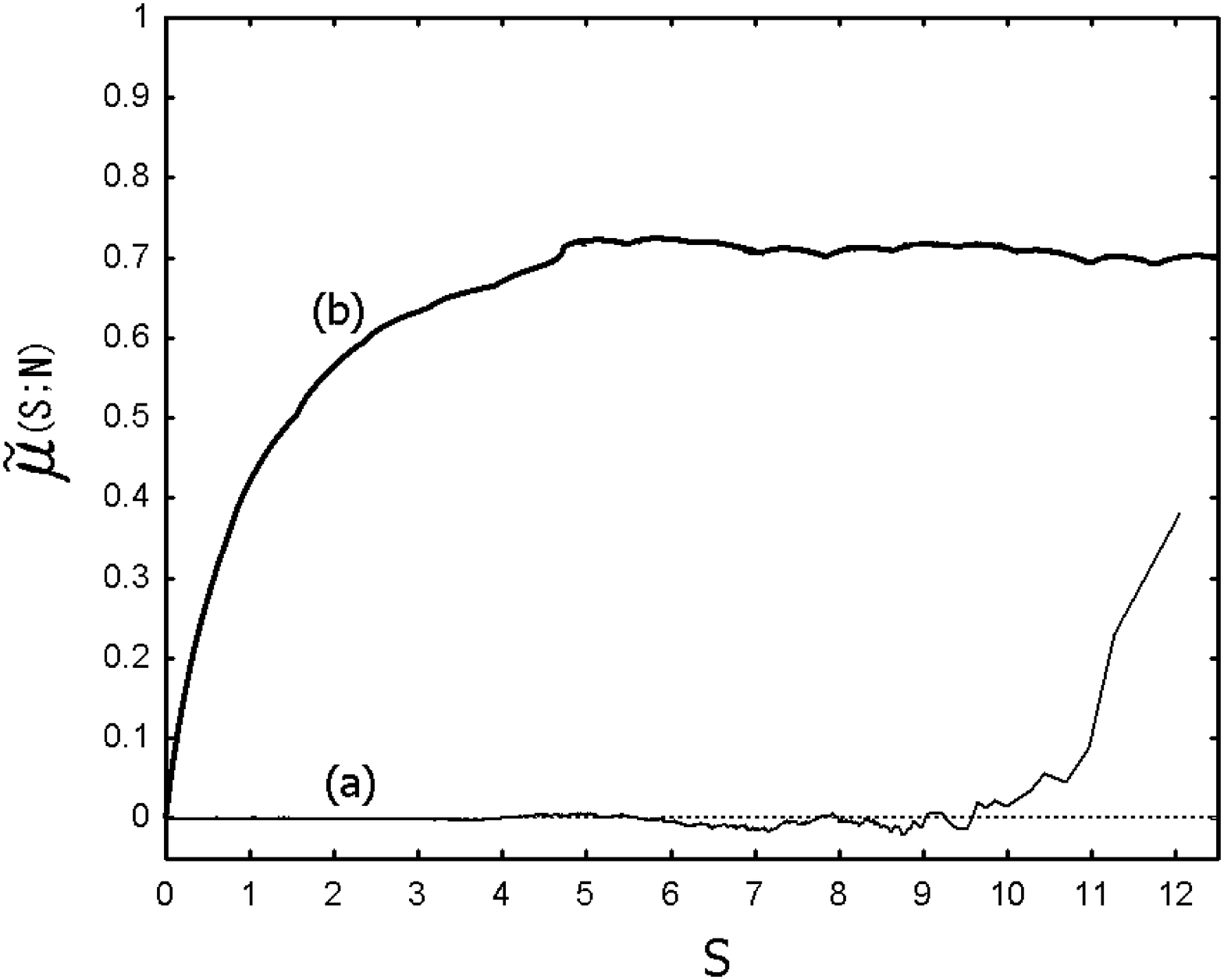}}
{FIG.4  The distribution function $\tilde{\mu}(S;N)$ for 
(a) $\alpha=1+\frac{\pi}{3}\times10^{-4}\ (N=61905)$, and 
(b) $\alpha=1+\frac{\pi}{2}\times10^{-9}\ (N=61906)$.  
The dashed line $(\tilde{\mu}=0)$ exhibits the Poisson 
distribution.}
\label{fig_3}
\end{figure}
\end{document}